\documentclass[12pt]{article}

\usepackage{graphicx}

\begin{document}

\begin{center}
{\Large\bf \boldmath Importance of fluctuations of cross sections \\
in muon-catalysed $t$-$t$ fusion reactions} %<== title (bold face, capitalize)

\vspace*{6mm}
{Sachie Kimura$^a$ and Aldo Bonasera$^{a,b}$ }\\      %<== authors
{\small \it $^a$ Laboratorio Nazionale del Sud, INFN,via Santa Sofia, 62, I-95123 Catania, Italy \\      %<== institutions
            $^b$ Libera Universit\`{a} Kore, via della Cittadella 1, I-94100 Enna, Italy}
\end{center}

\vspace*{6mm}

% abstract
\begin{abstract}
We discuss the reaction rate of the muon-catalysed $t$-$t$ fusion.
The reaction rate is determined as a function of the temperature using the model of 
``in flight" fusion.  
We especially take into account the effect of the fluctuation of the cross section 
due to the existence of the muon. The obtained reaction rate 
5.0$\times$10$^{-3} \mu$s$^{-1}$ is    
%1.1 $\mu$s$^{-1}$ in the low temperature region.
a factor of 10$^{-3}$ smaller than the experimental muonic cycling rate 
%This is comparable to 
%should be compared with 
%the muon cycling rate 
in the solid tritium target.
% 3.3$\pm$0.7 $\mu$s$^{-1}$.
%% is comparable to 
%% should be compared with 
%% the muon cycling rate in the solid tritium target, 3.3$\pm$0.7 $\mu$s$^{-1}$,

%% 1.1 $\mu$s$^{-1}$ at the low temperature is comparable to 
%% %should be compared with 
%% the experimentally obtained muon cycling rate in the solid tritium target, 3.3$\pm$0.7 $\mu$s$^{-1}$.
%% ***
\end{abstract}

\vspace*{6mm}

%% Use this template to prepare your manuscript for Proceedings of the
%% Conference.

%% Please, send your contribution to \verb|mcf07@theor.jinr.ru|. Deadline
%% for submission is September 30, 2007. Pagelimit is 8 pages.

\section{Introduction}

In the liquid hydrogen isotopes mixture, muons assist the fusion through the formation 
of a muonic molecule, since the size of the muonic molecule is much smaller than that of 
the ordinary molecules and the fusing nuclei tend to stay closer.  
%% After the fusion process the muon is released normally and again it is utilized for 
%% another fusion.   
This whole process takes place at the thermal energies where the conventional measurements
of the fusion cross section using a charged beam cannot be performed. 
This mechanism of the muon catalyzed fusion ($\mu$CF) might provide us an unique opportunity 
to investigate, in a rather direct way, the fusion cross section, i.e., the astrophysical
$S$-factor, at extremely low energy.   
%gives us informations has been studied as a realizable candidate of an energy source at thermal energies.
For this purpose we need to know quantitatively the effect that a muon shields the Coulomb 
potential between colliding nuclei~\cite{kb-amst}. 
And then this shielding effects of muons should be removed from the $S$-factor data, 
in order to asses the bare reaction rate correctly. From this point of view,
the $t$-$t \mu$ fusion could provide us an elucidating example. It 
%The muon catalyzed $t$-$t$ fusion 
has been investigated 
experimentally in the gas, liquid~\cite{breu} and solid~\cite{matsuzaki} targets.
%as a realizable candidate of an energy source at thermal energies. 
Especially in the latter experiments the fusion neutron energy spectrum has been determined and 
its distribution suggests 
%it has been clarified 
that the fusion is followed by a sequential decay of $^5$He: 
\begin{equation}
  \label{eq:ttann}
  t+t+\mu \rightarrow ^5{\bf He}^* +n \rightarrow \alpha + n+n + Q(11.33MeV),
\end{equation}
where $^5${\bf He}$^*$ is in the 3/2$^-$ and 1/2$^-$ resonant states. 
The $t$-$t$ fusion with muons has not 
received much attention as a candidate of an energy source 
%been studied so intensively 
in contrast to the $d$-$t$ and the $d$-$d$ $\mu$CF. This is partly because of the difficulty of tritium handling.
Moreover it is because its reaction rate is expected to be much lower than the others due to the lack
of the resonant muonic complex formation.  
In fact the cycling rate obtained experimentally: 3.3$\pm$0.7 $\mu$s$^{-1}$~\cite{matsuzaki} 
(15 $\mu$s$^{-1}$~\cite{petitjean,ackerbauer} ) of the $t$-$t\mu$ is 
smaller than that of the $d$-$t\mu$ of the order of 100.
Another distinctive difference of the $d$-$t~\mu$ and the $d$-$d~\mu$ reactions from 
the $t$-$t~\mu$ reaction is that their cycling rates has target density and temperature effects, 
which are likely caused by 3-body collisions. 
%% To the contrary, 
%% the reaction rate of the "in flight" fusion would not have density dependence beyond the first order, 
%% as it is clear from Eq.~(\ref{eq:rrvar}).
Put another way, if the cycling rate of the $t$-$t~\mu$ reaction does not have the target density and temperature
dependences, one can verify that the dependences originate from the formation of the resonant muonic complex.  

On the other hand, the cross section of 
the reaction $^3$H($^3$H,2$p$)$^4$He has been measured in the triton beam energy range 
$E_{lab}=$30-300~(keV)~\cite{tte-1,tte-2,tte-3,tte-4}. This energy range is much higher than 
thermal energies.   
The astrophysical $S$-factor of the reaction has been studied theoretically by means of 
DWBA~\cite{winkler} and the generator coordinate method~\cite{PhysRevC.50.2635}.  

We determine the reaction rate of the $t$-$t\mu$CF by considering so-called ``in flight" fusion~\cite{melezhik}
and compare it with the experimental muon cycling rate. 
%%  the framework of 
%% a semi-classical method, the constrained molecular dynamics (CoMD) approach~\cite{pmb}.
%***We determine the reaction rate as a function of the temperature. 
%Especially the screening effect by the bound muon is taken into account expricitly in the present paper.  ***
At thermal energies, where the $\mu$CF takes place, fluctuations of the cross section might play an important 
role. 
We investigate the influence of the fluctuations by using 
a semi-classical method, the constrained molecular dynamics (CoMD) approach~\cite{pmb,kb-ags}.
%As it is well known 
The molecular dynamics
contains all possible correlations and fluctuations due to the initial conditions(events). 
In the CoMD, the constraints restrict the phase space configuration of the 
muon to fulfill the Heisenberg uncertainty principle. 
The results are given as an average and a variance over ensembles of 
%the enhancement of the cross section 
the quantity of interest, which is determined in the simulation. 
%% Especially we determine the enhancement factor of the reaction cross section by the muon 
%% as a function of the incident energy in the ``in flight" fusion~\cite{melezhik}: 
%Especially 
We make use of the average and the variance of the enhancement factor of the 
cross section by the muon, that have been obtained in simulations of the $d$-$t \mu$CF~\cite{kb-amst},
and convert them into the average and the variance of the effective potential shift.   
If so, one can use the same potential shift for the case of the $t$-$t \mu$CF, because of the isotope 
independence of the screening effect. 
%By writing down the enhancement factor in terms of the potential shift, we  
We, thus, determine the reaction rate of the $t$-$t \mu$CF as a function of the temperature, 
taking into account the effect of the fluctuation of the cross section by the presence of the muon.
%on the reaction rate.

This paper is organized as follows.
%% In Sec.~\ref{sec:form} we describe the theoretical framework 
%% of the CoMD for muonic molecule formation and following fusion process briefly.  
In Sec.~\ref{sec:mrcs} we derive the variance of the effective potential shift 
of the reactions between hydrogen isotopes by the muons.
The reaction rate as a function of the temperature for the $t$-$t \mu$CF is determined 
in Sec.~\ref{sec:rr}. We discuss possible origins of the discrepancies between the experimental 
muon cycling rate and the obtained reaction rate in Sec. ~\ref{sec:rd}.
%The relation between the enhancement factor and the chaotic motion of the muon is discussed 
%% We develop in Sec.~\ref{sec:rad} a formula to estimate the initial $\alpha$-$\mu$ sticking 
%% probability($\omega_0$) and determine $\omega_0$.   
%Sec.~\ref{sec:strip} is devoted to the discussion of a possibility of muon release.  
In Sec.~\ref{sec:sum} we summarize the paper.
% and mention the future perspectives of this study. 

\section{Reaction cross sections in the presence of muons}
\label{sec:mrcs}
%Enhancement of the cross section by the muonic Screening effect}
%\label{sec:amsc}
The influence of the muonic degrees of freedom to the reaction cross section 
%of $tt$ 
can be taken into account as an enhancement by the screening
effect~\cite{kb-cdf, kb-icfe}.     
%The cross section is enhanced by the screening effect of the muon~\cite{kb-amst}.   
We determine the enhancement factor at the incident center-of-mass(c.m.) energy $E$ as 
%The enhancement can be written in terms of the potential shift $U_{\mu}$.
\begin{equation}
  f_{\mu}=\frac{\sigma(E)}{\sigma_0(E)},
\end{equation}
where $\sigma(E)$, and $\sigma_0(E)$ are the cross sections in the presence, and in the absence, respectively, of the muon.
The $\sigma(E)$ fluctuates depending on the dynamics of the 3(or N)-body system~\cite{kb-amst}.  
This fluctuation of the cross section can be written in terms of the fluctuation of the enhancement 
factor: $\Delta f_{\mu}$ as
\begin{equation}
  \Delta \sigma(E)= {\sigma_0(E)} \Delta f_{\mu},
\end{equation}
where we have assumed that the fluctuation of the bare cross section is negligible at low temperature
compared to the screened one. Through the molecular dynamics simulation we obtain this fluctuation 
as a variance of the enhancement factor. 
%% in the case of the muon catalysed $d$-$t$ reaction~\cite{kb-amst}. 
At the same time we determine the average enhancement factor: $\bar{f}_{\mu}$.
In our previous study we have simulated this enhancement of the cross section by the muon 
in the case of the $d$-$t$ $\mu$CF reaction~\cite{kb-amst}. We utilize the results from this simulation 
for the case of $t$-$t$ $\mu$CF reaction.   
%% From the constrained molecular dynamics simulation, 
%% we have obtained  
\begin{figure}[htbp]
  \centering
  \resizebox{70mm}{!}{\includegraphics{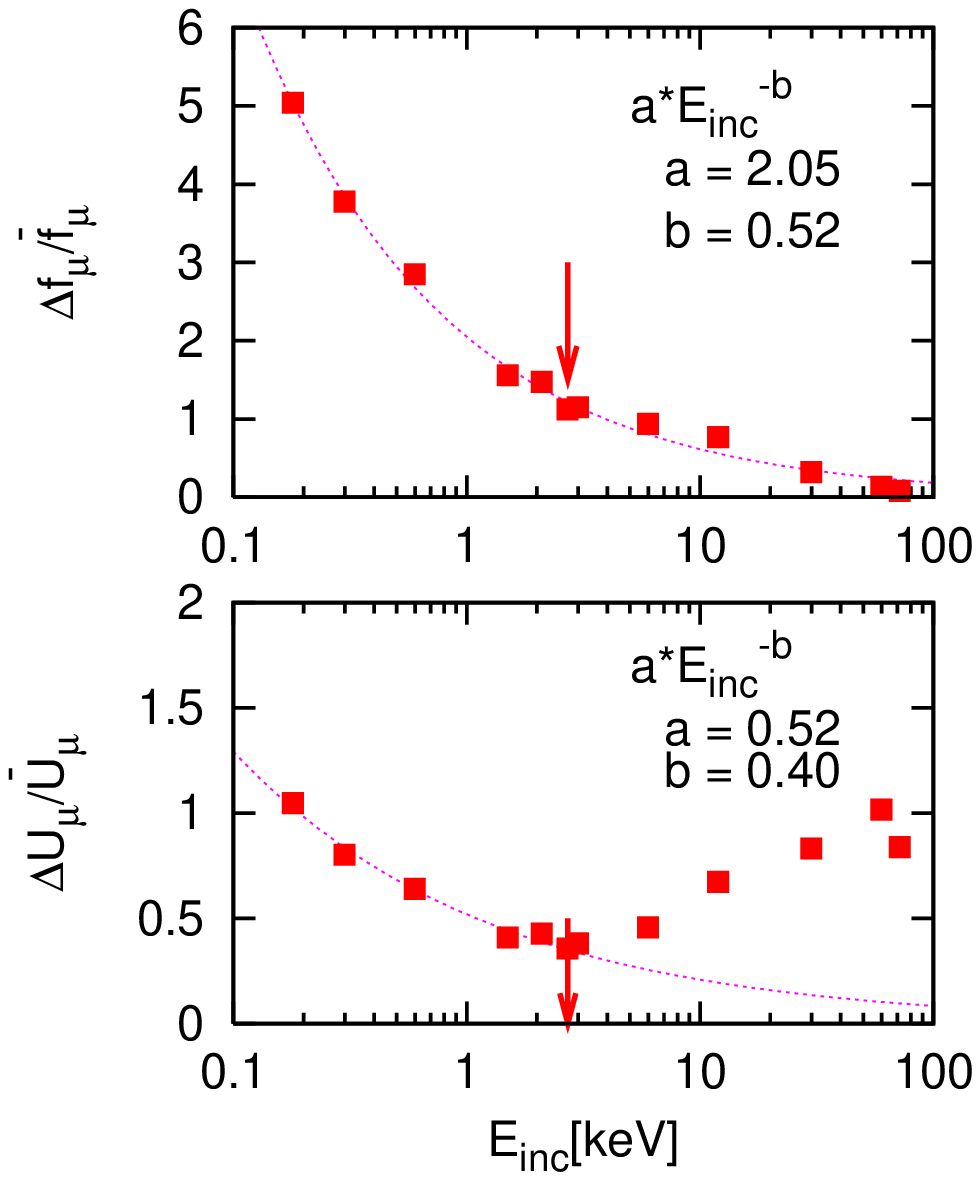}}
  \resizebox{70mm}{!}{\includegraphics{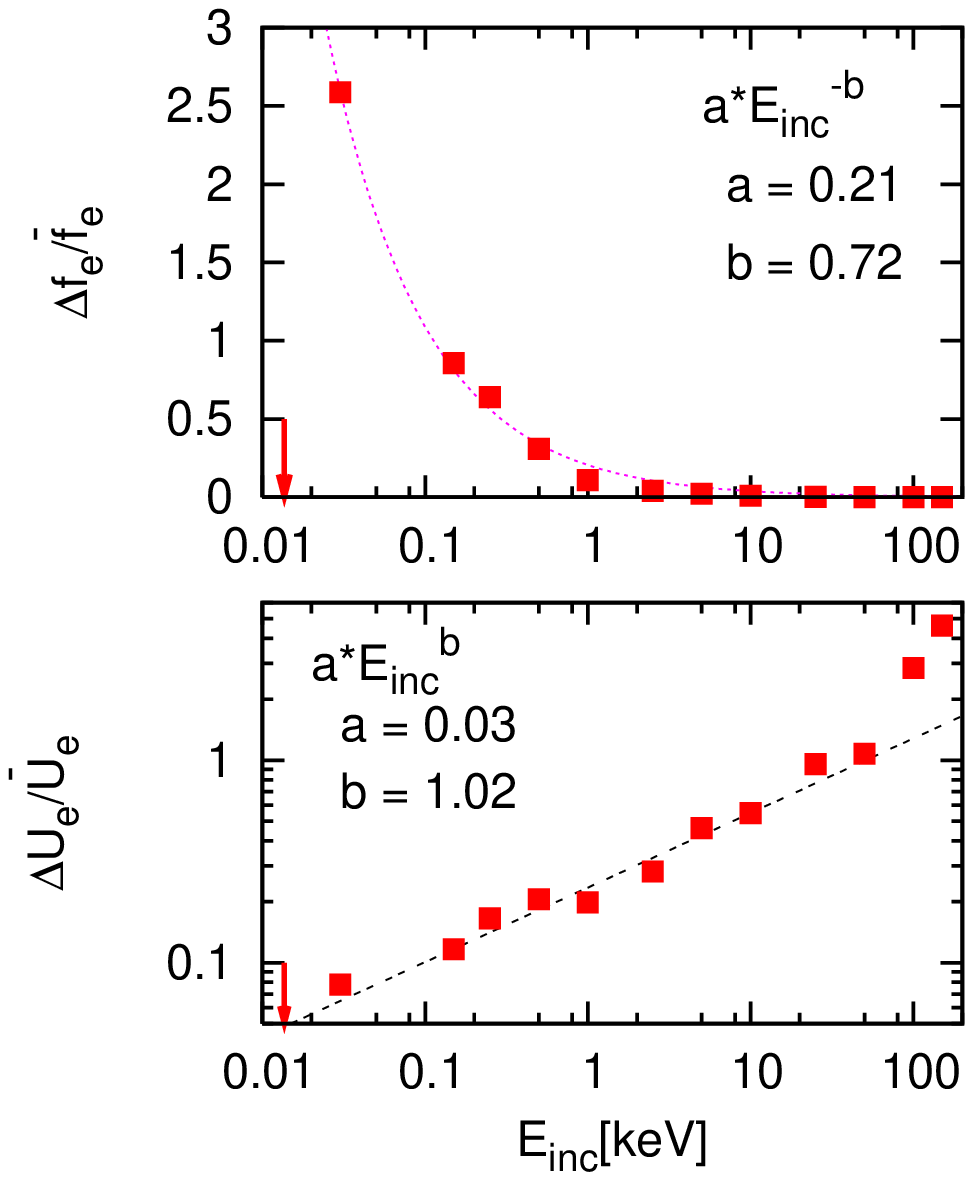}}
  \caption{(Left) The fluctuation of the enhancement divided by the average enhancement by the bound 
    muon~(the left-top panel)  in the $d$-$t\mu$ reaction.
    The fluctuation of the potential shift divided by the average potential shift~(the left-bottom panel), 
    both as functions of the incident c.m. energy.
    The arrows in the figure indicate the point where total energy is zero.
    The Right panels are same with left panels but for the $d$-$d$ reaction with a bound electron. }
  \label{fig:eu}
\end{figure}
In the left-top panel of
Fig.~\ref{fig:eu}, the ratio $\Delta f_{\mu}/\bar{f}_{\mu}$ is shown 
as a function of the incident c.m. energy. 
%% i.e., the fluctuation of the enhancement for each event in the simulation divided by 
%% the average enhancement by the bound muon 
%The feature of the figure can be explained 
In the high energy limit the ratio approaches zero, i.e., the $f_{\mu}$ distribution 
becomes a $\delta$-function ($\Delta f_{\mu}=0$) and the average ${f_\mu}$ approaches 1;
there is no effective enhancement. 
In the low energy limit the ratio $\Delta f_{\mu}/\bar{f}_{\mu}$ is much larger than 1; this fact
implies that the system exhibits 
a sensitive dependence of the dynamics on the initial conditions, i.e., the muonic motion becomes chaotic. 
The energy dependence of the ratio $\Delta f_{\mu}/\bar{f}_{\mu}$ is approximated well by 
the function 2.05$\times E_{inc}^{-0.52}$, where
$E_{inc}$ is in units of keV. This curve is shown by the dotted line in the left-top panel in 
Fig.~\ref{fig:eu}.

We write down the enhancement factor in terms of a constant shift of the potential barrier. Here we have 
assumed that the enhancement is represented in terms of a constant shift of the potential barrier
% assume that the enhancement is represented in terms of a constant shift of the potential barrier, 
%i.e., 
%the incident energy is shifted effectively from $E$ to $E+U_{\mu}$ 
due to the presence of the muon.
The average $\bar{f_{\mu}}$ for the $dt\mu$ is in good agreement with the exact adiabatic limit with 
the screening potential: 
\begin{equation}
U_{\mu}^{(AD)}=BE_t-BE_{^5{\bf He}} \sim 8.3 keV,
\end{equation}
where $BE_t$ and $BE_{^5{\bf He}}$ are the binding energies of muonic tritium and muonic $^5$He 
ion, respectively~\cite{kb-amst}. We, therefore, assume that the average potential 
shift $\bar{U}_{\mu}$ is equivalent to $U_{\mu}^{(AD)}$.  
From the commonly used expression of the bare cross section: 
\begin{equation}
\label{eq:sigma-s}
\sigma_0(E)=\frac{S(E)}{E}e^{-2\pi \eta(E)},
\end{equation}
where $S(E)$ and $\eta(E)$ are the astrophysical $S$-factor and Sommerfeld parameter.
The cross section in the presence of the muon is expressed by
\begin{equation}
\label{eq:sigma-ss}
\sigma(E)=\frac{S(E)}{E+U_{\mu}}e^{-2\pi \eta(E+U_{\mu})},
\end{equation}
in terms of the potential shift $U_{\mu}$.
Taking the derivative of the potential shift:
\begin{equation}
  \label{eq:dsigma}
  \frac{\Delta \sigma}{\Delta U_{\mu}}=\frac{\sigma}{E+U_{\mu}}\left(\pi\eta(E+U_{\mu}) -1 \right).
\end{equation}
%From now on 
Hereafter we substitute the potential shift $U_{\mu}$ in Eq.~(\ref{eq:dsigma}) by its average $\bar{U}_{\mu}$
and thus 
\begin{equation}
  \label{eq:dsigma2}
  \frac{\Delta \sigma}{\bar{\sigma}}=\frac{\Delta f_{\mu}}{\bar{f}_{\mu}}=\frac{\Delta U_{\mu}}{E+\bar{U}_{\mu}}\left(\pi\eta(E+\bar{U}_{\mu}) -1 \right).
\end{equation}
%by writing down the barrier penetration probability explicitly.  
One can deduce the average potential shift $\bar{U}_{\mu}$,
and its fluctuation $\Delta U_{\mu}$ from the ratio $\Delta f_{\mu}/\bar{f}_{\mu}$.  
\begin{equation}
  \label{eq:umu}
  \frac{\Delta U_{\mu}}{\bar{U}_{\mu}} = \frac{\Delta f_{\mu}}{\bar{f}_{\mu}}\frac{E+\bar{U}_{\mu}}{(\pi\eta(E+\bar{U}_{\mu})-1)\bar{U}_{\mu}}
\end{equation}
This potential shift is independent of the difference of isotopes, so that we make use of the 
same value in both cases of the $d$-$t$ and the $t$-$t~\mu$CF.  

In the left-bottom panel of Fig.~\ref{fig:eu} the fluctuation of the potential shift 
divided by the average potential shift is shown as a function of the incident energy.  
It is striking 
%noteworthy 
that the slope of the ratio $\Delta U_{\mu}/\bar{U}_{\mu}$ changes 
at the ionization energy of the muonic tritium, which we indicate by the arrow in the figure.  
At this incident energy the total energy of the system is zero.
The total system is unbound at the incident energies higher than this point, while 
the 3-body system might be bound at lower energies.
By contrast, as it is shown in the right-bottom panel in Fig.~\ref{fig:eu}, in the case of the bound 
electron screening the binding energy of the electron 
is much lower than the incident energy of our interest and the ratio $\Delta U_e/\bar{U}_e$ 
decrease monotonically as the incident energy decreases. Again the 
arrow indicates the ionization energy of the deuterium atom.          
%% \begin{figure}[htbp]
%%   \centering
%%   \resizebox{80mm}{!}{\includegraphics{plotEUel.ps}}
%%   \caption{Same with Fig.~\ref{fig:eu} but for the case of the $dd$ reaction with a bound electron. }
%%   \label{fig:euel}
%% \end{figure}
%%   \caption{Fluctuation of the enhancement divided by the average enhancement by the bound electron (top panel) 
%%      in the $Dd$ reaction.
%%      Fluctuation of the potential shift divided by the average potential shift.  
%%      (bottom panel) as functions of the incident center-of-mass energy.
%%    The arrows in the figure indicate the point where total energy is zero.}
The irregular muonic motion leads to smaller external classical turning point~\cite{kb-amst}. As a consequence 
the irregularity makes the enhancement factor larger opposed to the result of the electron 
screening~\cite{kb-cdf,kb-icfe}, where the irregular(chaotic) events give smaller enhancement factors. 
This contradiction is accounted for the fact that the system remains 
bound in the present case at low incident energies, while in the previous case even the lowest incident 
energy that was investigated is much higher than the binding energy of the electrons. 
Therefore the chaotic dynamics of the electrons causes to dissipate the kinetic energy between the target 
and the projectile and lowers the probability of fusion.    

%% \begin{figure}[htbp]
%%   \centering
%%   \resizebox{80mm}{!}{\includegraphics{plotEUobe.ps}}
%%   \caption{ }
%%   \label{fig:euobe}
%% \end{figure}

%% Once we obtain the average and variance of the potential shift, they are 
%% % This potential shift is 
%% independent of the difference of isotopes. 
%% %identical for different isotopes. 
%% Therefore we use the value    
%% The screening potential $U_{\mu}$ in the present case can be estimated using adiabatic 
%% approximation.~\cite{melezhik}
%$U_{\mu}=8.3$ keV.
%% In particular in the limit $E\ll U_{\mu}$ the fusion cross section can be 
%% written from eq.~(\ref{eq:defenh0}) 
%% and eq.(\ref{eq:cs}) as 
%% \begin{equation}
%%   \label{eq:cs2}
%%   \lim_{E\rightarrow 0}\sigma(E) = \lim_{E\rightarrow 0}\sigma_0(E+U_{\mu}) \rightarrow \frac{S(U_{\mu})}{U_{\mu}}e^{-2\pi \eta(U_{\mu})},
%% \end{equation}
%% where $\eta(U_{\mu})$ is the Sommerfeld parameter.

\section{Reaction rate}
\label{sec:rr}
In a liquid hydrogen tritium target at the temperature $T$, the velocity distribution, $\phi(v)$, 
of a pair of colliding particles is written as the Maxwellian distribution, 
\begin{equation}
  \label{eq:max}
  \Psi(E,T)dE= \phi(v,T)dv=\frac{2}{\sqrt{\pi}}\frac{E}{k_BT}e^{-\frac{E}{k_BT}}\frac{dE}{\sqrt{k_BTE}},
\end{equation}
where $E$ and $v$ are the relative energy and the velocity of the pair of colliding particles, in the present
case two tritons, and $k_B$ is Boltzmann constant.    
Although the experiment in~\cite{matsuzaki} has been performed using the solid target, 
in this paper we assume Eq.~(\ref{eq:max}) as the velocity distribution for simplicity. We will reconsider 
the validity of this assumption afterwards. 
The reaction rate 
%in a liquid hydrogen 
per pair of particles is given by~\cite{clayton}
\begin{equation}
  \label{eq:rr}
  <\sigma v> = \int \sigma(E)v \Psi(E,T)dE, 
\end{equation}
where $\sigma(E)$ is the reaction cross section in the presence of the muon. 
The reaction rate at the liquid hydrogen density $\rho_{LH}=4.25 \times 10^{22} cm^{-2}$:
\begin{equation}
  \label{eq:rr3}
  \lambda=\rho_{LH}<\sigma v>
\end{equation}
is obtained as a function of the temperature.
%is shown in the figure as a function of the temperature.

%***In the case of the solid target....

As we have seen in Sec.~\ref{sec:mrcs}, the cross section in the presence of the muon is expressed 
effectively with a potential shift $U_{\mu}$, hence we write down the reaction rate:
\begin{equation}
  \label{eq:rrtu}
  \lambda = \rho_{LH} \int \sigma_0(E+U_{\mu})v \Psi(E,T)dE, 
\end{equation}
where $\sigma_0(E+U_{\mu})$ fluctuates depending on the variance of $U_{\mu}$. 
This fluctuation of the screening potential can be incorporated through the following equation: 
\begin{eqnarray}
  \label{eq:rrvar}
  \lambda = \rho_{LH} \int \sigma_0(E+U_{\mu})v \Psi(E,T) N(U_{\mu}) dE dU_{\mu}, 
\end{eqnarray}
where $N(U_{\mu})$ is the distribution of the screening potential. As the first guess, the distribution is likely to 
be a normal~(Gaussian) distribution. In reality, however, we find that the distribution changes  
characteristically as a function of the incident energy. 
This change is observed at the point where the incident energy coincides with the ionization energy 
of the muon. 
\begin{figure}[htbp]
  \centering
  \resizebox{70mm}{!}{\includegraphics{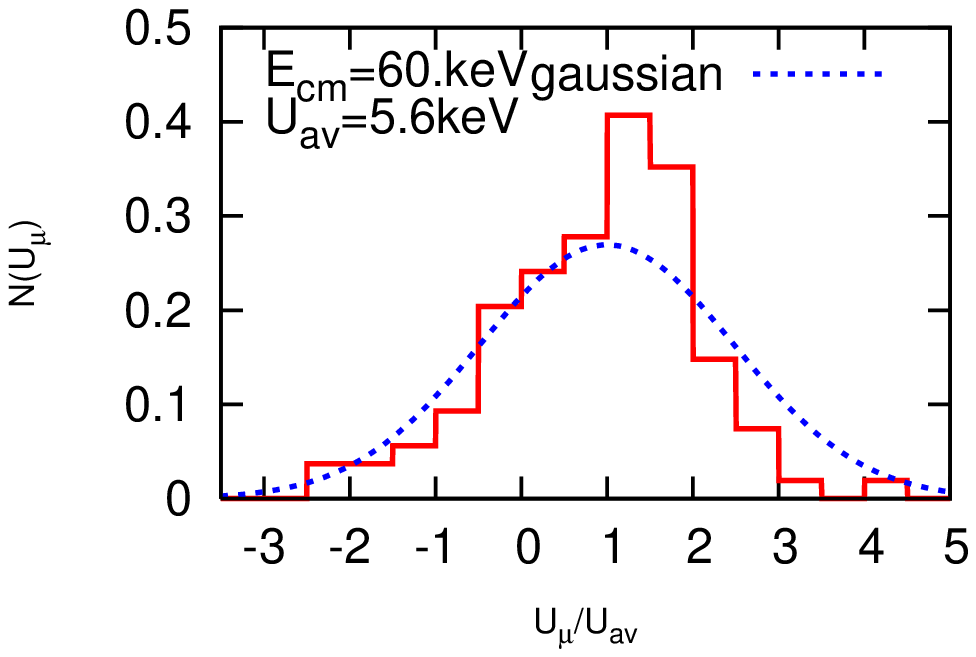}}
  \resizebox{70mm}{!}{\includegraphics{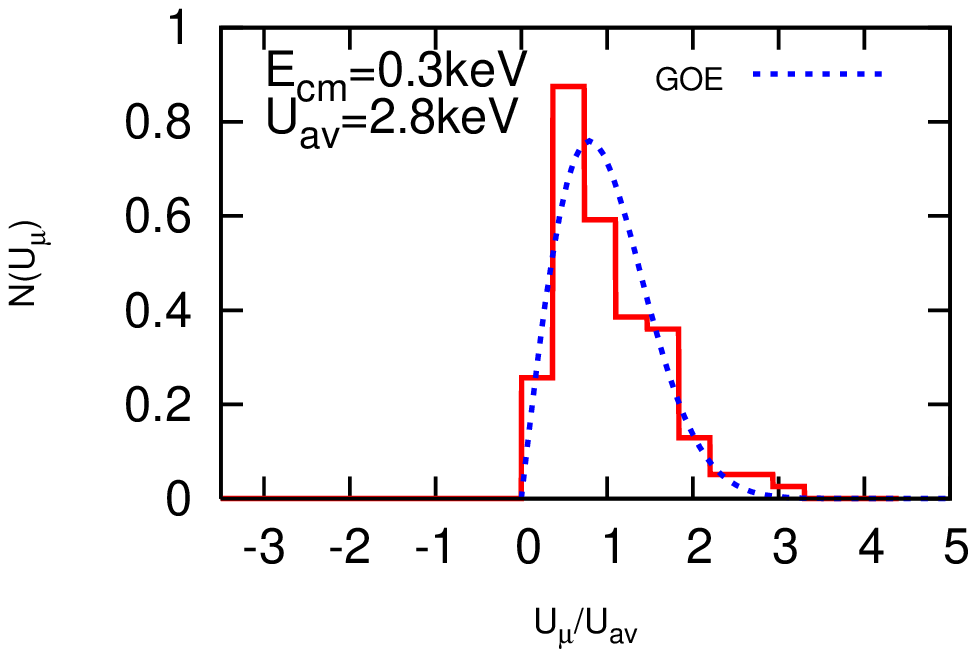}}
  \caption{ The histograms of the screening potential divided by the average at a relatively high incident energy~(the left panel) and at an incident energy lower than the ionization energy~(the right panel). }
  \label{fig:NNc}
\end{figure} 
In Fig.~\ref{fig:NNc} the left panel shows a histogram of the screening potential at a relatively high 
incident energy $E=$60~(keV). The abscissa is the screening potential normalized by the average value
$U_{\mu}=$5.6~(keV).  The distribution of the histogram is approximated well
by the Gaussian distribution, as we expected. Moreover one sees that there are some components with negative 
screening potentials. This negative component of the  
screening potential means that muons can be kicked out to an unbound state in some cases and bring away the 
relative energy of the colliding nuclei. 
%In another words, the muon brings the energy away from the relative motion. 
As the incident energy is reduced, the distribution changes. The right panel in Fig.~\ref{fig:NNc} shows one of such a situation: a histogram of the screening potential at the incident energy $E=$0.3~(keV).
In this case, 
%the case where the incident energy is lower than the ionization energy, 
because the muon is forced to remain bound the whole process in the entrance channel, 
the screening potential cannot be negative any more. 
In fact, we can approximate the distribution of
the histogram with the distribution of Gaussian Orthogonal Ensemble~(GOE) rather than 
a normal distribution. 
We, therefore, approximate this fluctuation of the screening potential by Gaussian distribution around 
the average potential shift $\bar{U}_{\mu}$ and with variance $\Delta U_{\mu}$: 
\begin{equation}
 N(U_{\mu})=\frac{1}{\sqrt{2\pi}\Delta U_{\mu}}\exp\left(-\frac{(U_{\mu}-\bar{U}_{\mu})^2}{2\Delta U_{\mu}^2}\right). 
\end{equation}
in the energy region $E> BE_t$ and by GOE: 
\begin{equation}
 N(U_{\mu})=\frac{\pi}{2}\left(\frac{U_{\mu}}{\bar{U}_{\mu}} \right) \exp\left(-\frac{\pi}{4}\left(\frac{U_{\mu}}{\bar{U}_{\mu}}\right)^2\right) 
\end{equation}
in the energy region $ E < BE_t$.
%% Thus we obtain the reaction rate:
%% \begin{eqnarray}
%%   \label{eq:rrvar}
%%   \lambda = \rho_{LH} \int \sigma_0(E+U_{\mu})v \Psi(E,T) N(U_{\mu}) dE dU_{\mu}, 
%% \end{eqnarray}
Using these distributions of the screening potential, we assess the Eq.~(\ref{eq:rrvar}). 
%In Eq.~(\ref{eq:rrvar})
We substitute the bare cross section by Eq.~(\ref{eq:sigma-s}) and 
we use the polynomial expression~\cite{winkler}
\begin{equation}
  \label{eq:sfact}
  S(E)=0.20-0.32E+0.476E^2~({\rm MeVb}),    
\end{equation}
where the energy $E$ in MeV,  
as the astrophysical $S$-factor
%bare reaction cross section $\sigma_0(U_{\mu})$ 
in Eq.~(\ref{eq:sigma-s}).
%in the NACRE~\cite{nacre} 

\section{Results and discussions} 
\label{sec:rd}
\begin{figure}[htbp]
  \centering
%  \resizebox{70mm}{!}{\includegraphics{plotRtt.ps}}
  \resizebox{70mm}{!}{\includegraphics{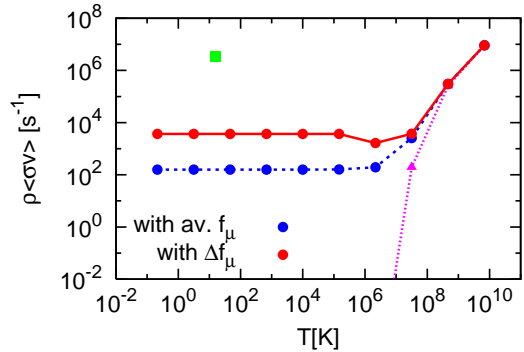}}
  \caption{Reaction rate of the $tt\mu$ system at the liquid hydrogen density as a function of the target temperature.
  The circles linked with the dotted line are obtained taking into account 
  the average enhancement of the cross section by the muon, alone.
  The circles linked with the solid line are obtained taking into account 
  the fluctuation of the enhancement of the cross section by the muon.
  The triangles show the bare reaction rate. 
  The square shows the experimental data of the muon cycling rate. }
  \label{fig:Rtt}
\end{figure}

The obtained reaction rate by Eq.~(\ref{eq:rrvar}) is shown in Fig.~\ref{fig:Rtt}
% shows the reaction rate 
by circles linked with a solid line as a function of the temperature. 
%at the liquid hydrogen density($\rho_{LH}=4.25\times10^{22}cm^{-3}$).
The circles linked with the dotted line are obtained by taking into account 
the enhancement of the cross section using the average potential shift $\bar{U}_{\mu}$ alone, i.e., 
Eq.~(\ref{eq:rrtu}) substituted $U_{\mu}$ by $\bar{U}_{\mu}$. 
%The filled squares linked with the solid line are from Eq.~(\ref{eq:rrvar}) considering the fluctuations 
%of the potential shift, as well. 
The triangles show the bare reaction rate.
The circles deviate from Eq.~(\ref{eq:rrtu}) at the temperature lower than 10$^8$ K and 
do not show much temperature dependence from 0.1 K to 10$^5$ K.
% however, , which is relevant to the experimental data. 
The deviation of the reaction rate by Eq.~(\ref{eq:rrvar}) 
from the one by Eq.~(\ref{eq:rrtu})
%the case where only the average enhancement is considered 
%the former 
in the low temperature region indicates 
that the fluctuation of the enhancement factor has a strong influence on the reaction rate
at the thermal energy.          
%The latter deviate from the former 
Including the fluctuation of the cross section,  
the reaction rate for the $t$-$t \mu$CF reaches at 5.0$\times$ 10$^{-3} \mu$s$^{-1}$.   
%1.1 $\mu$s$^{-1}$ in the low temperature region.
This is 
a factor of 10$^{-3}$ smaller than the experimental muonic cycling rate 
%This is comparable to 
%should be compared with 
%the muon cycling rate 
in the solid tritium target, 3.3$\pm$0.7 $\mu$s$^{-1}$~\cite{matsuzaki}, which is marked by the square
in Fig.~\ref{fig:Rtt}.
% and smaller than the mean half-life of the muon.  
We remark several possible explanations of this underestimation of the rate in the following.
The first one is that the assumption of the velocity distribution of colliding nuclei to be the  
Maxwellian could be inappropriate. We should make use of the velocity distribution in the solid target.    
Otherwise the second possible explanation is that assuming that the procedure of the estimation of the reaction 
rate is correct, 
the result could mean that the actual bare $S$-factor must be greater than that given by 
Eq.~(\ref{eq:sfact}) in the low energy region where the measurement has not been performed.

%What one can deduce from this result is that 

%% One can give an appropriate estimate of the reaction rate of $t$-$t~\mu$ reaction 
%% in the framework of the "in flight" fusion. 
%% This is because the ``in flight'' reaction is a major contribution to the muon cycling rate in the 
%% case of the $t$-$t~\mu$CF, while in the case of the $d$-$t$ and the $d$-$d~\mu$CF the muonic molecular complex  
%% %the muonic molecule 
%% formation rate controls the muon cycling rate.    

%% Last, another distinctive feature of the $d$-$t~\mu$ and the $d$-$d~\mu$ reactions from 
%% the $t$-$t~\mu$ reaction is that their cycling rates has target density effects, which 
%% are likely caused by 3-body collisions. 
%% To the contrary, 
%% the reaction rate of the "in flight" fusion would not have density dependence beyond the first order, 
%% as it is clear from Eq.~(\ref{eq:rrvar}). 

\section{Conclusions and Future Perspectives} 
\label{sec:sum}
 
In this paper we have estimated the reaction rate 
in the muon-catalysed $t$-$t$ fusion as a function of the temperature
in the framework of the "in flight" fusion. 
We have used the CoMD simulations to estimate the enhancement effect of the cross section due to the muon.
We found that the estimated reaction rate does not show temperature dependence in 
the low temperature region in contrast to the experimental data of the $d$-$t$ and the $d$-$d \mu$CF.      
The obtained reaction rate in the low temperature: 
5.0$\times$ 10$^{-3} \mu$s$^{-1}$ underestimates    
%1.1 $\mu$s$^{-1}$ in the low temperature region.
the experimental muonic cycling rate 
%This is comparable to 
%should be compared with 
%the muon cycling rate 
in the solid tritium target, 3.3$\pm$0.7 $\mu$s$^{-1}$, a factor of 10$^{-3}$. 
%% is comparable to 
%% should be compared with 
%% the muon cycling rate in the solid tritium target, 3.3$\pm$0.7 $\mu$s$^{-1}$,
This is either because our assumption that the velocity distribution is Maxwellian  
is not correct in the case with the solid target or 
because the $S$-factor, which we used to estimate the rate, in the low energy region
is lower than the actual value.

\end{document}